\documentclass{kluwer}  
\usepackage{epsfig}

\newdisplay{guess}{Conjecture}
\begin{document}                                                                                   
\begin{article}
\begin{opening}         
\title{A Review of Disk-Corona Oscillations} 
\author{Jean \surname{Swank}}  
\runningauthor{J. Swank}
\runningtitle{Disk-Corona Oscillations}
\institute{NASA/Goddard Space Flight Center}

\begin{abstract}
Low frequency ($\approx 0.1-35$ Hz) quasi-periodic oscillations of the X-ray
flux characterize many of the black hole candidates, in particular
those which have radio evidence for jets. These QPO have amplitudes of
up to 20 \% in the states of black hole novae which are called very
high and intermediate and are correlated with the break frequency of
the band-limited low frequency white noise. With transition to the
state called the soft high state, both the QPO and the noise
disappear.  While the noise is strong in the low hard states like that
of Cyg X-1, the QPO, when present, are weak and broad. In their strong
manifestations, these QPO have the curious property of appearing to
have the spectrum of the power-law component which dominates in the
low state, while correlations between their frequency and the disk
component in the spectrum imply control by the disk. The correlations,
the harmonic structure of the QPO, and the phase lags have complex
behavior in the same source (GRS 1915+105, XTE J1550-564).  The
phenomena point to interaction between the disk and the corona,  
for which there are several interesting ideas.


\end{abstract}
\keywords{quasi-periodic oscillations, microquasars, black hole candidates }

\end{opening}           

\section{Introduction}  
Disk-corona oscillations, as discussed in this paper, 
refer to the oscillations in X-ray
flux, appearing as quasi-periodic oscillations (QPO) in the power
density spectra, which are in the non-thermal spectra possibly
originating in a corona, but which appear to be related to
optically-thick emission associated with the accretion disk. The
QPO can have an amplitude of 10--20 \%. 
In the power
density spectra of black hole candidates (BHCs), a subset of which are known
to be microquasars, several low frequency features have appeared. For
a given source several QPOs can appear simultaneously, in which case it is
clear that there are more than one phenomenon manifested. 
The characteristics of the features 
seen in the range 1--15 Hz are best known, but they have been seen below
1 Hz and above 15 Hz.

These oscillations were seen with {\em Ginga} in GX~339--4 (Miyamoto et
al. 1991) and Nova Muscae (Takizawa et al. 1997). The 
frequencies increased with the source intensity, not always
in the same way.
When {\em RXTE} discovered GRS~1915+105 to be active in 1996, the
oscillations were soon found to be prominent during long episodes of
hard flux and relative quiescence, and the oscillation frequencies
were found to be strongly correlated with the source intensity (Chen
et al. 1997). 
GRO~J1655--40
became active and showed the same kind of features (Remillard
et al. 1999).
By the end of 1999, {\em RXTE} had observed recurrences of the black hole
candidate transient 4U~1630--47, and 4 new transients, all of which
exhibited at some stages very high amplitude low frequency QPOs: 
GRS~1739--278,
XTE~J1748--288, XTE~J1550--564, and XTE~J1859+226. 

Cyg X--1,  1E~1740.7--2942,
GRS~1758--258, and several transient sources which did not get very bright,
have been known for band limited white
noise. Sometimes there have been  peaks and even
a harmonic in the frequency range where the spectra roll over
(Cui et al. 1997; Smith et al. 1997). These have  lower
amplitudes, 5 \%, and sometimes are hard to distinguish from a sharp
transition in the power density spectrum.
In Cyg~X--1 the amplitude
was highest during the transitions between  low and high states, 
appropriately called the ``intermediate state''.

\section{Characteristics of the QPO}

Usually the QPO feature has at least one harmonic. Sometimes a peak at
what would be the subharmonic is present. For simplicity we refer to
the dominant peak as representing the phenomenon, whatever may be the
real fundamental of the oscillation. 

Almost always it is present over a range of luminosities and the
frequency varies with luminosity. In most cases the frequency has
increased with count rate and indeed with luminosity, when the spectrum
has been taken into account. GRO~J1655--40 was a notable exception,
with the frequency generally decreasing with the total luminosity. 
However, Sobczak et al. (2000) found that frequency was 
positively correlated with the flux of the soft spectral component.
There may be some counter-examples. In 4U~1630--47, a
square wave oscillation of flux at a lower frequency was observed
(Dieters et al. 2000) and two low frequency QPOs were observed in the 1-10
Hz range, one moving with the flux, the other stationary. 
These frequencies are at least a factor of 10 too low to be related to
the inner most stable orbit of the accretion disk, for which the Kepler 
frequency would be 2200 Hz $(M/{M_\odot})^{-1}$. 

The energy spectra of the sources can usually be well or almost
described by an optically thick component and a non-thermal
component. 
A physical model would be a disk spectrum Comptonized in a
hot corona, but it can be approximated  by the
a multicolored disk and a power-law. 
The QPO usually appears when the flux is
dominated by a relatively hard  power-law-like component.

The energy spectrum of the QPO fundamental is that of the power-law  
component. Cui et al. (1999) plot (for
XTE~J1550--564) the root mean square (rms) amplitude of the QPO
fundamental and harmonic  with energy. The
fundamental tracks the energy spectrum. Curiously the harmonic
reverses and declines with energy, in comparison.

It was shown (Wijnands \& van der Klis 1999b) that for a number of
observations of BHCs, the break frequency of the band
limited white noise component was about a fifth of the QPO
frequency. However Morgan et al. (1997) exhibit power density spectra with
the QPO frequency clearly corresponding to the break frequency. The
exact connection between the two
requires further study. 

For GRS~1915+105 there is a great deal of data from different states
and correlations between characteristics have been studied with
several approaches. Markwardt et al. (1999) looked at many
observations in which on time scales of minutes, the flux varied
strongly (factors of 2-10). They averaged power spectra for given flux
levels and found that the average exhibited a strong correlation of
frequency of the QPO with the bolometric flux of the disk component in
a spectral fit (4 s spectra), above the frequency of about 4
Hz. 
Muno et al. (1999) computed spectral
parameters and power spectra for 16 s intervals of a larger set of
observations, including those in which variations were slow. Frequency
appears correlated with disk inner temperature for many of
these. Plots with respect to disk black body flux, temperature, and
inner radius all show runs of correlated data points, but also subsets
clearly violating the correlation or having a very different
correlation, indicating the behavior is not unique.

It is tempting to think of the frequency as related to the Kepler
frequency at some radius, although the factor and physical origin of
the relation may not be clear. Trudolyubov et al.
(2000) leapfrog from the dependences of $ f_K $ and the viscous time
scale at the radius to the relation between the duration of the hard
state and the minimum frequency observed. The
duration and frequency are easy to measure. The inner radius of the
disk is difficult to measure, especially when the disk inner
temperature appears to be low and the radius large, so that the {\em RXTE}
PCA can only ``see'' the Wien tail of the energy spectrum.
This gives
$t \propto f_K^{-7/3}$ with a proportionality depending on the rate of
mass flow through the shell. For the sets of observations with hard
dips, this correlation appears to be a good fit. For the observations
with soft dips, the dependence is significantly steeper.

For a given mass accretion rate in the disk, the temperature has a
constrained relation to the radius, of course. Belloni et al. (2000) 
examined whether this relation holds and allows identification
of the mass flow in the disk. Their aim was to compare to the mass
loss rate in the radio emitting plasma that is ejected. 
These questions are just the sort that can
give physical insight into the correct description of the disk
accretion and the generation of winds and jets, but the X-ray 
calibration to date limits the conclusions.

The confusion of opposite dependence between frequency and total count
rate in GRO~J1655--40 and GRS~1915+105 appeared clarified by their
having similar correlation with the disk flux
alone. Furthermore, Sobczak et al. (2000) found that for
XTE~J1550--564, while the dependence of the QPO frequency with either
temperature or radius was complex, the frequency did increase  with
disk flux. But from
the point of view of the Accretion Ejection Instability (Tagger \& Pellat 
1999; Rodriguez et al. 2000) the complex
dependence of the QPO frequency on radius is just what would be
expected if the inner disk of GRO~J1655--40 is nearer to the black
hole, where relativistic effects can cause a turnover of the curve. 

To interpret these correlations, physical models are needed for which
the data can determine the parameters reliably. In deducing radii of
the disk, even with the caveat that it is the optically thick disk, we
are using zero-order approximations,  when we
know that we should consider the scattering atmosphere of the disk 
together with advective and wind flows
(Merloni et
al. 2000). However, although the meaning of the slopes may need modification,
the fits to the data show that correlated changes occur, with 
values of the parameters often in realistic ranges.

\section{Relation of Disk-Corona Oscillations to other QPO}
Although much of the discussion  has been in
terms of the inner radius of an optically thick accretion disk that is
being seen in the X-ray spectrum, it should be remembered that the QPO has
appeared to occur simultaneously with the QPO that appears at much higher
frequencies. It has seemed reasonable
that the latter could reflect either the Kepler orbital frequency
near the Innermost Stable Circular Orbit
about black holes or the eigenfrequency of the fundamental modes of
oscillation of the inner disk. 
High frequencies  have been reported
for 5 BHCs. 
For GRO~J1655--40, XTE~J1550--564, and  XTE~J1859+226,
they have occurred together. 
Obviously if the higher frequency reflects emission from near the inner edge
of the disk, the lower frequency cannot be the same thing, although it 
could be a  different time scale at the same place. Alternatively,  the
inner disk may not be otherwise seen directly, while the lower
frequency could reflect a transition region in the disk 
rather than the inner radius.


For GX 339--4 and GRS 1739--278 the low frequencies were only seen
around 6 Hz and 5 Hz respectively. For Nova Muscae, 
GRS~1915+105, 
XTE~J1550--564, 4U~1630--472, XTE~J1859-277, values sampled include  
similar ranges within 0.08-18 Hz,
while the high frequencies of the last 4 of these were 67 Hz, 161-237 Hz, 
185 Hz, and 150-200 Hz. For two of the sources, GRO~J1655--40 and 
XTE~J1748--288, the low frequencies were higher, 14-35 Hz.
GRO~J1655--40's high frequency was also higher, 300-400 Hz. 
Dynamical masses 
are so far known only for
GRO~J1655--40 and Nova Muscae.
How the frequencies scale with black hole  mass is not yet clear.

QPOs at frequencies an order of magnitude below these 
disk-corona oscillations also occur. One type of variation which
generates very low frequency oscillations in the power spectra are
square-wave modulations of the flux originally called ``flip-flops''
by Miyamoto et al. (1991), and also dips. 
GRS~1915+105 has a 15 mHz feature whose folded light curve looks almost 
like an
occultation (Morgan et al. 1997). 

The disk-corona oscillation is in a frequency range just below the
Horizontal Branch Oscillations (HBO) of Low Mass Binary Neutron Stars
and they can look similar in harmonic structure.
Usually the
amplitude is lower. 
The HBO and Normal Branch frequencies of the Atoll and Z sources
have been found to be correlated to
the lower of two kilohertz frequencies.  
For the BHCs, Psaltis et al. (1999) see
features, sometimes very broad, that they consider to be counterparts of the
neutron star lower kilohertz frequency. 
They find the disk-corona oscillation and this 
oscillation to be related in the same way as are the HBO and the 
lower kilohertz oscillation. The ratio of the
average lower frequency in the BHCs and of the HBO in the neutron star
sources could be consistent with being inversely proportional to the
mass of the compact object 
(Chen et al. 1997). 

\section{Relation to Black Hole Accretion States}
Most black hole candidate transients have outbursts which last several
months, but not as long as has GRS~1915+105. They usually have a well
defined onset, rise, peak, and decay, although there are rather
frequently one or more brightenings during the decay. GRS~1915+105 has
for several years now evolved from one recurrence to
another. GRO~J1655--40 
and XTE~J1550--560 recurred after a year, but they had clearly gone into
quiescence first. In a  typical case  
there is a rising phase, frequently a very high
state (VHS), a high soft state (HS) and decay through an intermediate
state (IS) to a low hard state (LS) and finally a quiescent state. 

{\em RXTE} has had a number of successful observations of rising phases
(XTE~J1550--564, Cui et al. 1999; 4U~1630--472, Dieters et al. 2000;
XTE~J1859+226, Markwardt et al. 1999) In these cases the spectra started out
harder than achieved subsequently in the outburst. The spectrum
softens as the flux rises. The disk-corona oscillation starts below 1
Hz with high amplitude and rises in frequency, increasing in width and
decreasing in amplitude somewhat. 
But the time scale for these changes is
days rather than the minutes required for the GRS~1915+105 hard dip recoveries.

At the end of the rising phase, the transient may be in the VHS with a
strong power law and a strong disk flux. 
In both the
VHS and IS the QPO is present. In the HS the soft component is very
bright (several tenths of the Eddington limit). A
relatively steep power-law component extending to high energy
may be of a completely different origin than the Compton up scattering
in a hot corona that is favored for explaining the VHS. There are NO
QPOs in the HS.


In the case of GRS1915+105, we discuss a scenario of mass flow in the
disk, removal of the disk into a corona and ejection into jets. Does
this scenario apply to the transient outbursts?
In the VHS
there is outflow, because the radio emission often starts sometime during
the rising phase. The mass accretion rate in the outer disk is large
and the inner radius of the disk is large. As the mass flow rate
subsides the disk collapses to be optically thick, or at least it is
optically thick farther in toward the center. 
At a lower rate, there are also radio injections associated with the
transitions to the HS and out of it to the IS
(e.g. Hjellming et al. 2000) from comparison of the radio and  
X-ray transitions (for
XTE~J1748--288 and XTE~J1859+226). 
This scenario would imply that spectral fits obtain larger inner
radii for the disk in the VHS than in the HS, but only a factor of 2
seems to have been observed. Much needs to be done to understand the evolution.

\section{Explanations for the Low Frequency QPO}
The flux that is oscillating appears to be the hard power-law component of the
spectrum, the candidate for indicating the presence of a
hot corona. The thermal photons from the disk would be  the dominant 
source of the
seed photons to be up-scattered in the corona. 
Whether the corona is spherical and central or
extended over the inner disk, or whether there is simultaneously
spherical infall is not clear. Radio observations are indicating 
(See Fender, this volume)
outflow and synchrotron radiation. 
A number of models have been suggested and would explain some of the
observations.  I 
only mention some of the ideas.

Disk-corona oscillations are probably not like the 
possible coronal radial infall
oscillations suggested for the normal branch oscillations at about 6
Hz in the neutron star Z sources. 
Neither observations or simulations show
the large amplitude and variability.

It has been suggested that the HBO in neutron stars are due to the
Lense-Thirring frame dragging in the strong gravitational field close
to rather heavy neutron stars (twice the nodal 
precession frquency of slightly tilted and
elliptical orbits.) 
Systematic similarity of the 
neutron stars and BHCs has been claimed, as
mentioned above.  In some
cases the approximately linear relation between two frequencies in the
BHCs and the HBO and lower kilohertz frequencies for the neutron stars
is impressive.
Psaltis \& Norman (2000) have made a case that the frequencies
observed are the resonant frequencies in the disk, whatever the mechanism
for modulating the X-ray flux. 

One idea that appears to generate the right order of magnitude of
frequencies and similar dependences to those observed is that of
shocks at the inner disk and oscillation of the shock position 
(Molteni et al. 1996; Manickam \&
Chakrabarti 1999), but there have been objections that 
such shocks cannot be generated.  The
correlations between duration of the dips in GRS 1915+105 with the
frequency of the QPO are difficult to distinguish from the
correlations expected in a viscous refilling model (Trudolyubov et
al. 2000). 

Finally, Tagger \& Pellat (1999) make a
case that Rossby waves of the Accretion Injection Instability are a
natural way to carry gas and energy from a disk to the corona and
thence into a wind or jet and to generate oscillations in the flux.

The data available so far has enough detail to  provide 
many guideposts and constraints to the theories. These
low frequency disk-corona oscillations are
very common in BHCs and are a challenging physics problem which
involves  important aspects of accretion onto black holes.

\end{article}
\end{document}